# Spherical Indexing for Neighborhood Queries

Nicolas Brodu

*Abstract*—This is an algorithm for finding neighbors when the objects can freely move and have no predefined position. The query consists in finding neighbors for a center location and a given radius. Space is discretized in cubic cells. This algorithm introduces a direct spherical indexing that gives the list of all cells making up the query sphere, for any radius and any center location. It can additionally take in account both cyclic and non-cyclic regions of interest. Finding only the K nearest neighbors naturally benefits from the spherical indexing by minimally running through the sphere from center to edge, and reducing the maximum distance when K neighbors have been found.

## I. Introduction

Finding the neighbors of a given point is a general problem that has attracted much intention [1] and that is still a major topic of research. Some solutions are more appropriate than others for specific applications, especially when handling large dimensional data, or when the objects are static and their position can be pre-sorted. The algorithm presented here concentrates on dynamic objects, for which the position cannot be known beforehand and changes with time.

The case for three dimensions is described in this article, but the algorithm can be generalized to other dimensions as well. It can be seen as an improvement over the bin-lattice spatial subdivision method [2], itself a particular bucketing algorithm.

The neighborhood query problem consists in finding the objects within a given radius from a given center location, either all of them or only the K nearest. This defines what is called the *query sphere* in this document. The naive algorithm to answer the neighborhood query is O(n): run through the list of all objects and compute their distance, then compare with the query radius to find the neighbors. Unfortunately when the query is repeated for each object, for example in the case of a multi-agent simulation where each agent wants to find its neighbors, then the naive algorithm becomes $O(n^2)$ and doesn't scale. For large simulations, neighborhood queries quickly dominate the computing costs, and the program spends more time resolving such requests than doing the actual simulation work with the result of these requests!

Let's now consider as in Fig. 1 a discretization of space consisting in a regular lattice of cubic cells. With the extra assumption that objects are represented by their position in space, each point object will be assigned to one cell, and only one. Each cell may contain as many as all the objects, or it can be empty. The idea behind the bin-lattice spatial subdivision method is to quickly eliminate all cells that are beyond range, outside the sphere, with the consequence that all objects within these cells are eliminated without computing their distance to the query center. Only the cells near the query center need to be processed. With the bin-lattice method, cells beyond the N1 norm cube (see Fig. 1) are not considered.

But the query sphere volume is $4/3 \cdot \pi \cdot r^3$, and its bounding cube volume is $2 \cdot r^3$, so the sphere fills only about 52% of the cube. This ratio is also the limit of the number of cells intersecting the sphere over the number of cells within the cube as the discretization size tends to 0. For a large distance query with respect to the cell discretization, up to nearly half the cells could thus be rejected. For higher dimensions, these sphere / cube volume ratios tend to decrease quickly, from about 31% in four dimensions to less than 1% for dimensions 9 and above [3].

The idea with the spherical indexing method introduced in this document is to not even consider the extra cube cells that do not intersect the query sphere. The assumption is that the cost to set up the spherical indexing is lower than the cost of considering cells outside the sphere, so there will be a final gain. Section IV investigates how well this assumption holds in practice. The algorithm costs may be classified into 2 broad categories:

- Processing the cell. This is where the distance comparison is effective, as well as other checks like emptiness, that allow to potentially reject a cell.

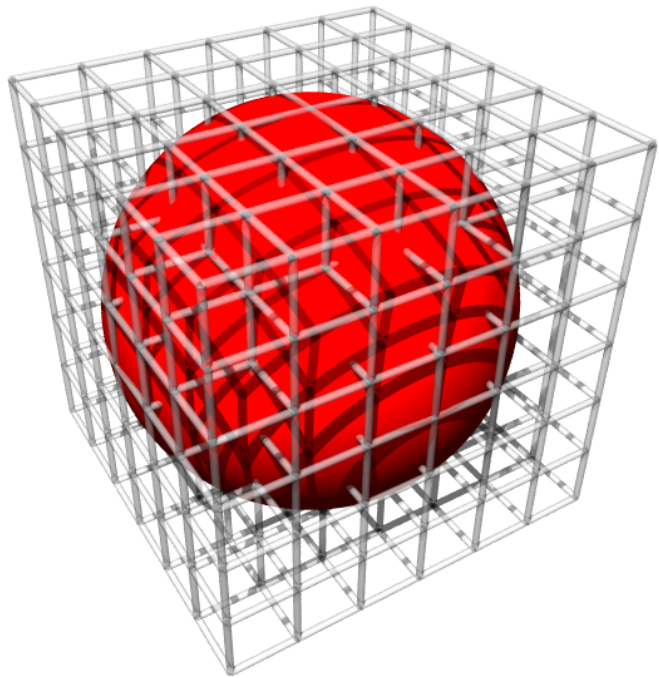

Fig. 1. Query sphere intersection with a discretized space: The center is the location for which neighbor objects should be found within a certain radius. Some cells in the cube corners do not intersect the query sphere and should not be considered.

Nicolas Brodu is a PhD student at the Department of Computer Science and Software Engineering, Concordia University, Montreal, Quebec, Canada, H3G 1M8 (e-mail: nicolas.brodu@free.fr)

This work was financed in part by the EADS Corporate Research Center, with the support of the French Ministry of Foreign Affairs.

- Accessing the cell. At least one memory line must be loaded to process anything on a cell, including a test for emptiness or a distance rejection. Since the number of cells and the size of the data structure is quite large, it cannot fit all in cache. The access cost is therefore dependent on the number of cells accessed, whatever their content, and bounded by the memory speed, not the CPU speed .

For light load situations, where there are few objects per cell, saving on the second kind of cost becomes important. For heavy load situations, with many objects per cell, saving on the first kind of cost is important.

The core of the algorithm presented in this document is an indexing scheme that gives the list of all cells intersecting the query sphere for any radius and any center location. This is done very efficiently by means of a few bit masking and shifting operations, and a single table lookup that specifies the list of all cells to process, implemented as a precomputed sorted array. The cells that are not in this list are not subject the the aforementioned costs.

The next section presents the spherical indexing scheme and the cells representation. Section III introduces important optimizations, especially for small query distances, comparatively to the discretization size. Section IV extensively analyzes the behavior of the algorithm and identifies the influential factors for best performances. Section V presents possible extensions. The conclusion in section VI discusses some applications of this algorithm. A reference implementation is introduced in Appendix.

## II. SPHERICAL INDEXING SCHEME

### A. Indexing cells by distance

Let's note the query center C, the query radius d, and the distance between C and an object X by x. The objects X in neighborhood are thus those for which $x \leq d$.

With these notations, the rejected cells are *at least* strictly d away from C, with the distance between C and a cell defined as the minimum distance between C and the points in that cell. The other cells are either intersecting the sphere boundary or completely inside it.

The first step consists in building a lookup table based on the minimum distance two points in different cells can be. This table is precomputed only once at program startup, or it may be loaded from an external file.

To build the table, let's consider an arbitrary cell as the starting point. The question of translating this arbitrary cell to the query center will be addressed in I.B. Let's assume for now this cell contains the query center location.

Fig. 2 shows the two-dimensional table lookup building process. The minimum bound for the squared distance from the query center to the points in a given cell is computed. Points in the center cell may potentially be at the same location as the query center, so the minimum bound is 0. Points on the cells surrounding the center are at minimum 1 floating-point unit of least precision distance, but mathematically the minimum bound for these cells is 0 too. Points in cells further apart get increasing minimum bounds, expressed in this document in cell units. The three and upper-dimensional versions of this process can be deduced from this

| 18 | 13 | 10 | 9 | 9 | 9 | 10 | 13 | 18 |
|---|---|---|---|---|---|---|---|---|
| 13 | 8 | 5 | 4 | 4 | 4 | 5 | 8 | 13 |
| 10 | 5 | 2 | 1 | 1 | 1 | 2 | 5 | 10 |
| 9 | 4 | 1 | 0 | 0 | 0 | 1 | 4 | 9 |
| 9 | 4 | 1 | 0 | 0 | 0 | 1 | 4 | 9 |
| 9 | 4 | 1 | 0 | 0 | 0 | 1 | 4 | 9 |
| 10 | 5 | 2 | 1 | 1 | 1 | 2 | 5 | 10 |
| 13 | 8 | 5 | 4 | 4 | 4 | 5 | 8 | 13 |
| 18 | 13 | 10 | 9 | 9 | 9 | 10 | 13 | 18 |

Fig. 2. Minimum bound for the squared distance between any two points respectively inside the center cell and a given cell, in two dimensions. The distance is noted in this document in cell units. An example is provided for the computation of the cells with gray background: $3^2+2^2=13$ and $2^2+1^2=5$.

Fig. 2 example. In three dimensions there are 27 cells at distance 0, 54 at $d^2=1$, 36 at $d^2=2$, 8 at $d^2=3$, 54 at $d^2=4$, and so on.

Some squared distances are missing, for example 7. These distances can be reached by truncating a squared query radius with floating-point computations, but by construction they do not bring in any new cells. Such distances are therefore treated as the non-empty entry just below them.

Cells are then represented by their offset in each dimension from the center cell, as will be detailed in the next section I.B. The list of all offsets for each non-empty squared distance minimum bound is maintained.

The lookup table is at this point an array of offset lists, indexed by the squared distances. Given a target query radius d, there is an integer $n = \lfloor d^2 \rfloor$ = floor($d^2$), such that $n \leq d^2 <$ n+1. As aforementioned, all cells that are at least strictly d away from the center query are rejected. Therefore, all cell offsets at n+1 and above are rejected. Consequently, it is sufficient to truncate $d^2$ so as to get the table lookup entry corresponding to that distance: all cells strictly below n+1 are these below or equal to n by construction.

The next step is to organize the cell offsets for a $\lfloor d^2 \rfloor$ table entry contiguously in memory just after the non-empty entry just below it. Thanks to this layout, the table lookup will need only be used once in run-time: to give the last distance, where to stop in this global offset array. The list of all cell offsets inside the sphere is then returned from center to edge contiguously, sorted by increasing distance.

This is particularly useful for K-nearest neighbor queries. When the user is interested in finding only the K nearest neighbors, it is possible to adjust the maximum distance based on the $K^{th}$ found neighbor, as soon as K neighbors are found: potentially closer candidates are necessarily within radius equal to the current furthest found neighbor distance.

### B. Representing cells by their offsets

The space discretization is assumed to be finite, defined over a region of interest. This section relies on a power-of-two sized discretization in each dimension, for maximal performance. Non-power-of-two sizes could be implemented by extension, but this limitation is usually acceptable, and well worth the optimization it brings.

Thanks to the power-of-two assumption, each cell can be given an absolute linear index within the region of interest, corresponding to its binary representation. As an example, let's consider sizes of respectively 32, 16 and 8 in X, Y, and Z. A cell at position 22 in X, 10 in Y and 3 in Z would be given the absolute index (in binary): `011_1010_10110`, in ZYX order and with underscores added for clarity. This linear index is also the position in memory of that cell in a large array containing all the cells in the region of interest. This index is called the *packed location* of the cell in this document.

An *unpacked location* format is also introduced. It corresponds to shifting the Y component so as to ease operations further on. The same example would be unpacked as `1010_00000_011_0000_10110`. Note that zeros were inserted, and the Y component was shifted to the left by the total number of bits in the packed format. The unpacked index can be shifted back to the right `1010_00000` and OR'd with a masked version of itself `011_0000_10110` to get the original packed index.

As described in the previous section, each cell in the precomputed sphere composition array is represented by its offset from the center cell. These offsets are written in unpacked format.

This algorithm can handle both cyclic and non-cyclic worlds, in any direction. An example is given first for the cyclic case, the non-cyclic case is deduced by extension.

For completely cyclic worlds, the offset is written directly in 2-complement in each dimension. An offset of -5X, +4Y and +3Z would be written in unpacked format as `0100_00000_011_0000_11011`. The leftmost bit of each component is also the bit sign in 2-complement arithmetic (X is -5, not +27).

Given such an offset and a cell center, both in unpacked format, it is easy to find the final absolute position of the cell for that offset:
- Add both numbers. Re-using the same examples:
  `1010_00000_011_0000_10110` (center)
  + `0100_00000_011_0000_11011` (offset)
  = `1110_00000_110_0001_10001` (position)
- Mask out bits from 2-complement overflow
  & `1111_00000_111_0000_11111` (mask)
  = `1110_00000_110_0000_10001` (final pos)
- Pack the final position `110_1110_10001`

This is a form of parallelism, similar to [4], that effectively processes all three dimensions in a single register. This way, the sphere is translated to any center cell in the region of interest. In this example, the cell at offset -5X, +4Y and +3Z from the center at 22X, 10Y and 3Z, was correctly packed into the cell at 17X, 14Y, 6Z.

In the case of non-cyclic directions, the offsets are actually written in 2-complement on n+1 bits for each dimension, with n the number of bits in that dimension. This allows to represent both positive and negative offsets with full range on the n bits. The same masking as above gets rid of the extra bits. However, care is taken to also map cells outside the region of interest into a unique index `1_000_0000_00000`, so as to preserve the linear nature of the large memory array containing all the cells: A unique cell is added for all objects outside the region of interest. The reader is invited to consult the reference implementation for more details.

To sum up, to find all cells within a given locality sphere for any given center and radius, the operations are a table lookup, directly running up the array of cell offsets from center to edge, and for each cell a few bit shift/mask and addition operations as described in this section.

## III. OPTIMIZATIONS

### A. "Shaving" the sphere extrema

The full benefit of the algorithm comes with avoiding the consideration of cells in the query sphere bounding cube, but not intersecting the sphere. But as aforementioned in the introduction, the sphere volume over its bounding cube volume ratio tends to ~52% only as a limit case, when the discretization size is small compared to the query size. Actually, by analogy with Fig. 1, so long as the bounding cube corner cells are not entirely outside the query sphere, the sphere offset method cannot reject them, hence cannot bring any advantage. Moreover, for small distances, the pre-computations introduced in Fig. 2 may actually bring in more cells than necessary. For example, when the query radius is less than the cell size, the bin-lattice algorithm may consider between 1 and 27 cells, whereas the sphere offset method would at this point always consider the 27 cells by construction.

But in the case of Fig. 1, a visual inspection shows that 32 cells are completely outside the sphere, over 216 cells, hence a gain of 14.8%. That gain is not negligible, and even if it is not the theoretical maximum of about 48%, the spherical indexing method needs to handle that situation.

The solution to this problem is to consider the sub-cell location of the query center. In Fig. 2, if the query center is on the right side of the middle cell (grayed 0), then the cell immediately on its right may perhaps be at distance 0, but not the cell immediately on its left: the minimum distance depends on the location of the query center within its cell. Fig 3. shows more details about what happens in each dimension.

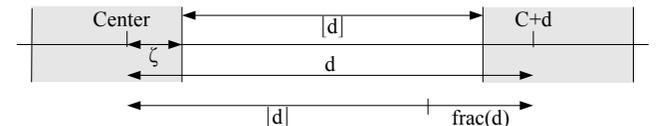

Fig. 3. Some relevant distances between two grayed cells. The center cell is on the left, the target cell on the right. d is the query distance. $\lfloor d \rfloor$ = floor(d) is the largest integer below or equal to d. It is also the distance between the two cells. frac(d) is $d - \lfloor d \rfloor$, the fractional part of the distance. $\zeta$ is the distance from the center to the cell edge in the direction of the target cell.

The target cell, on the right, should be rejected if $C + d < \lceil C \rceil + \lfloor d \rfloor$, where $\lceil C \rceil$ = ceil(C) is the lowest integer above or equal to C. In that case, the C + d would be closer than the target cell closest edge. Thus, the target cell may be safely omitted if $d - \lfloor d \rfloor < \lceil C \rceil - C$, or in other words, frac(d) < $\zeta$.

This result is generalizable to each of the 6 directions, counting the three axis positively and negatively. In the negative case, the target cell would be on the left in Fig. 3, and then $\zeta$ = frac(C).

Rejecting cells this way is actually equivalent to what the bin-lattice algorithm does. The problem mentioned in the introduction to this section is then solved: the sphere indexing algorithm now considers only the same cells out of the 27 as would the bin-lattice algorithm. But it is not yet solved for higher distances. Moreover, we can do better, with a precomputed test and a run-time test that are introduced below. These tests finally complete the initial goal: the sphere algorithm considers less cells than the bin-lattice one, even for d<1.

Cells that are diagonally placed from the center have not yet been considered. For example, in Fig. 2, when the query distance is 2, the frac(d) < $\zeta_i$ criterion would successfully eliminate the cells that are labelled "4", in the directions where this is possible. However, the cells that are labelled "2" are already $\sqrt{2}$ away from the center cell, and may very well be at d > 2 away considering the query center sub-cell location.

Fig 4. shows the situation similar to Fig. 3 in two dimensions. The base distance of a target cell is b, with $b^2$ an integer that is also the table entry for the sphere indexing. Let's note t the true distance from the query center C to the cell. The cell can be rejected if t > d, or equivalently:

$t^2 > d^2$, since both are positive

$\sum_{i=x,y} (b_i + \zeta_i)^2 > d^2$, with $b_i$ and $\zeta_i$ positive distances

$\sum_{i=x,y} (b_i + \zeta_i)^2 > (\lfloor d \rfloor + f)^2$, with $f = frac(d)$

$b^2 + \zeta^2 + 2 \sum_{i=x,y} b_i \zeta_i > \lfloor d \rfloor^2 + f^2 + 2 \lfloor d \rfloor f$    Eq. 1.

The Fig 4. diagram and these formula can easily be extended to higher dimensions. The next step is to find conditions for the cell rejection that can be precomputed.

Re-using the previous condition for the rejection in one dimension, let's assume that $f < \zeta_i$ for each direction i. Let's additionally assume that $b \geq \lfloor d \rfloor$. Then:

$b^2 + \zeta^2 + 2 \sum_{i=x,y} b_i \zeta_i > \lfloor d \rfloor^2 + 2 f^2 + 2 f \sum_{i=x,y} b_i$    Eq. 2.

by direct application of the assumptions in two dimensions. But, thanks to the triangular relation, $b_x + b_y \geq b$, and with the previous assumption, $b_x + b_y \geq \lfloor d \rfloor$. Since $f \geq 0$ by definition, Eq. 2 shows that the set of chosen assumptions satisfies Eq. 1 and the cell can be rejected.

This result is especially interesting because it can be precomputed. In each of the 6 directions, there is the possibility that $f < \zeta_i$ or not. This gives $2^6 = 64$ combinations, leading to as many specialized offset tables where the cells satisfying Eq. 1 are not included. These tables should only be used for $b \geq \lfloor d \rfloor$, but that's easy to ensure: the main table is run from sphere center to edge in increasing distance. The specialized tables are used as soon as $\lfloor d \rfloor$ is reached and no sooner. The precomputation relies on the fact that since the tables are only used for entries $b^2$ such that $d \geq b \geq \lfloor d \rfloor$, then by definition of $\lfloor d \rfloor$ = floor(d), $\lfloor d \rfloor = \lfloor \sqrt{b^2} \rfloor$ for each table entry $b^2$: $\lfloor d \rfloor$ is known at precomputation time even if d is not.

This optimization is particularly useful because it only has a low run-time cost, related to selecting the right specialized table out of the 64. Once this is done, all cells that are pre-eliminated do not even need to be brought in memory and then tested for potential rejection. This is a net gain.

However, the chosen set of assumptions is itself not optimal: some cells may satisfy Eq. 1 but not Eq. 2. These cells will not be pre-excluded, but could still be rejected at run-time, at the cost of an additional check.

Fortunately, thanks to the offset representation, all the $b_i$ distances in Eq. 1 are quickly available. It is then just a matter of adding the $\zeta_i$ and testing for $t^2 > d^2$ to decide whether to reject the cell or not. Cells with $b < d - \sqrt{3}$ are below one cube diagonal of the maximum distance. They need not be tested as they will always be included. In practice the run-time check takes time, and it was observed that it is not worth its cost if applied unconditionally for all cells above $d - \sqrt{3}$. Since cells below d - 1 are also always included along the 6 main directions, it was decided to apply the run-time test only for cells above d - 1. Extensive testing has confirmed that d - 1 is a good heuristic for applying the run-time test: Some cells in diagonal configurations escape from rejection, but the test is not uselessly applied to the cells at d-1 in each direction.

Then, as for the bin-lattice algorithm, objects in the cells that are still present so far are individually tested for rejection. This induces a cost that is proportional to the cell load: the average number of objects present in each cell. A final optimization is to unconditionally include all objects for cells below $d - \sqrt{3}$. Indeed, in that case, the cells are entirely within the query sphere, and so are the objects within these cells.

### B. Using the exact number of processed cells information

The distance table building process described in section II implies that cell offsets are sorted by increasing distance order from the query center. The total number of cells that will be processed for a given radius can then be estimated in advance: these are the cells that fall partially or completely within the query sphere. Too far away cells are by construction excluded from this list, but cells on the edge are conditionally included or not depending on the center sub-cell location, as discussed in the previous section. The solution is to pre-compute for each distance the probability of rejecting the cells not entirely within the sphere, by performing a sampling of the possible center sub-cell locations. The average number of cells for a given distance is now known, and can be used to make a decision at run-time.

The main algorithm runs through the cell list in query sphere range, and rejects empty cells to speed up the neighbor search. A completely different algorithm exhibits the opposite of that situation. The idea is to combine them so as to get the best performance.

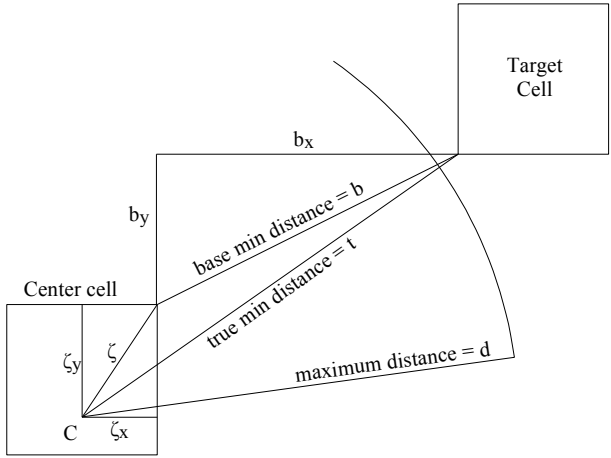

Fig. 4. Sphere shaving process: Considering the query center position inside the center cell allows to reject cells at maximum distance when $b \leq d < t$.

Whenever an object position changes in the region of interest, it is both easy and efficient to maintain a linked list of non-empty cells. If the algorithm was running through the non-empty cells list instead of the sphere cells list, it would not need to check for empty cells by definition, but would rather need to check for cells that are too far. The worse case would be one object per cell, in which case running the non-empty cells list degenerates to the naive neighborhood query algorithm. The best case would be immediate rejection because all objects are in a single far away cell.

Choosing the non-empty cells list versus the sphere cells list is thus a matter of trading off rejection because a cell is too far, by rejection because a cell is empty. Disregarding the difference in the processing costs of these checks, the decision is a simple matter of choosing the list with the smallest number of elements. By definition, the cells that are both non-empty and within query radius are included in both lists, so the smallest list will bring in less rejections than the other.

The non-empty cell list is being used only when the query distance is quite large and the cell load average quite small. In this case there are few non-empty cells, less than the sphere volume. In practice, experiments suggest that such a situation is undesirable anyway. It is often better to reduce the discretization of the region of interest so there is a smaller total number of cells, and a higher load ratio. Nevertheless, maintaining the non-empty cell list has a negligible run-time cost. This potential optimization has been kept in the reference implementation due to the benefits it brings when it is applicable.

*C. Large distance optimization*

Boundary phenomena cannot be neglected when the query distance becomes large.

For a non-cyclic region of interest the offsets must be considered up to the maximum distance, in both the positive and the negative directions. So, the sphere diameter can be as large as twice the size of the region of interest. In that case, the sphere volume may also be greater than the total world volume.

Even if it is not, the sphere may cover only a small part of the region of interest compared to the sphere volume. For example, when the query center is located in a corner of the region of interest, only one eighth of the sphere is inside. The outside cells are all mapped as described in II.B. to a unique cell, so their memory access cost is reduced as the unique outside cell is already in cache after the first use. Nevertheless, accessing unused outside cells induces as many costly checks for rejection.

A solution would be to parametrize the sphere clipping according to the distances to the region of interest boundaries, and adapt the offset tables. However, the sheer amount of combinations make this solution intractable in terms of memory consumption.

A sub-optimal solution is to consider the intersection of the sphere bounding cube with the region of interest. This defines a parallelepiped of cells that minimally covers the portion of the sphere inside the main zone, equivalent to the region the bin-lattice algorithm would consider.

The volume of that parallelepiped then becomes another quantity, that can be checked against both the sphere volume and the non-empty cell list size. Then, three comparisons are enough to select the most appropriate method at run-time.

In the case of a cyclic world in every dimension, the table offset building method makes the sphere cover more and more of the world finite volume, up to the point where it covers all the region of interest. For cyclic worlds there is no boundary problem: clipping the query distance to the maximum is enough to both cover the whole region and to avoid duplicates.

## IV. BENCHMARKS AND INFLUENTIAL FACTORS

*A. Finding all neighbors*

All the plots presented in this section were obtained according to the same method. Each measurement consists in 150 randomly centered queries that were repeated 40 times, and the 10 worse results were discarded. The justification is to avoid occasional spurious jitter due to the interfering multi-tasking. The remaining 30 measurements were then averaged to further damp out the system fluctuations. The functor applied to each neighbor accesses the memory line of the neighbor object, so as to realistically simulate a simple operation.

The distance is increased by 0.1 cell size steps until an arbitrary point of ¾ the region of interest size is reached. No qualitative change is observed beyond that point, and usually even before, especially for wrapping worlds.

The bin-lattice neighborhood query algorithm was reimplemented with the same data structures as the proposed algorithm for fairness of comparison. A previous and publicly available bin-lattice algorithm [5] using simple double-linked lists is also shown in Fig. 7 to demonstrate the importance of the data structures and the implementation.

Fig. 5 shows the relative performance improvement of the proposed algorithm over the bin-lattice one in a cyclic world. A ratio of X means that the proposed algorithm performs X times faster than the bin-lattice algorithm. The result is a neat improvement over a wide range of distances, with high ratios in some cases.

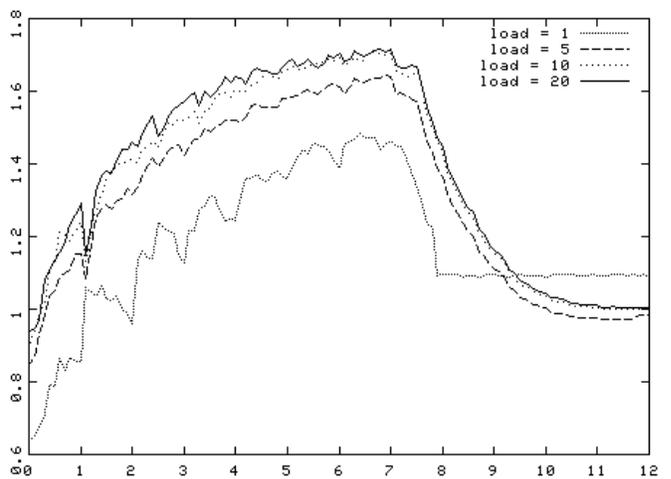

Fig. 5. Proposed algorithm / bin-lattice performance ratio vs query distance (cell units), for a 16x16x16 wrapping world, for different cell load averages measured in objects/cell.

The gain clearly depends on the average number of objects per cell, a direct consequence of not considering cells the bin-lattice would consider. When the load average is too low, the cells the spherical algorithm avoids are mostly empty, so there isn't much gain. When the load average becomes high, so does the gain of avoiding the processing of a cell.

For short distances below 1 cell unit, the bin-lattice algorithm has the advantage of simplicity. For short distances, the setup costs for the more complex algorithm can only be compensated by eliminating diagonal cells that the bin-lattice cannot, when these cells processing cost justifies it. This is the case for high load averages.

For distances in a cyclic world greater than half the world size minus half a cell size, the bin-lattice algorithm covers the whole world, and a simple optimization can be made: The unconditional inclusion of all objects in the region of interest! This explains the sharp drop observed for such distances above 7.5. The query spheres with increasing radius can only unconditionally include objects in cells entirely within the sphere: this explains the result for the case of large distances with a load average 5 in Fig. 5, where the bin-lattice algorithm performs slightly better (ratio close but lower than one).

For higher load averages, the sphere method still considers less cells than the whole world, and these savings in processing costs result in better performances. However for low load averages the non-empty list becomes interesting: it contains less cells than the sphere volume. The proposed algorithm switch to using the non-empty list is clearly visible for the load average of 1 in Fig. 5.

Fig. 6 shows the absolute performances for that situation. The performances are plotted on a logarithmic scale so as to offer a global view of the algorithm response to different query distances. Both the bin-lattice and the purely spherical algorithms are also shown, for comparison. Neither can switch to the non-empty list, but the bin-lattice method uses the aforementioned whole world optimization. All three effects are visible on the graphic: the sphere progressive saturation, the non-empty cells list usage, and the whole world consideration.

Fig. 7 is the analogous situation to Fig. 6 in a non-wrapping world, with a higher load average of 10. The

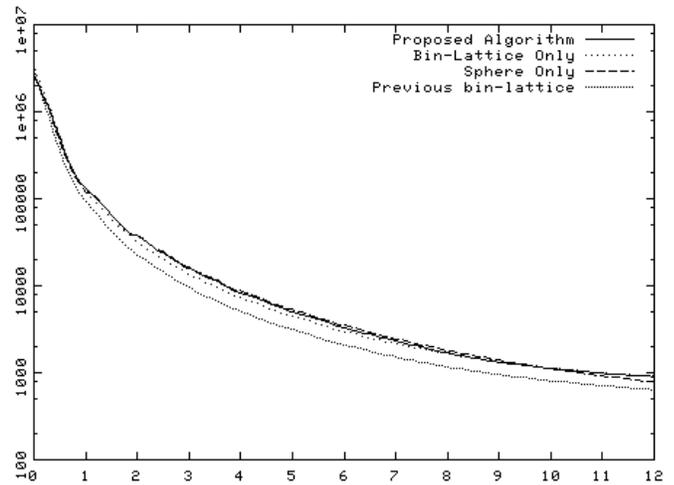
Fig. 6. Performances (queries/s) vs query distance (cell units), for finding all neighbors within range, in a 16x16x16 non-wrapping world, with a load average of 10 objects/cell.

aforementioned previous bin-lattice implementation using double-linked lists is presented as well.

Comparing the two bin-lattice curves shows the optimization only due to the implementation and data structures. This implementation improvement was found to depend mainly on the load average, with lower loads exhibiting the best ratios. The proposed algorithm curve can be compared against the bin-lattice one using the same data structures, so as to determine the algorithmic improvement independently of the implementation. This was done in Fig. 5, and it is repeated in Fig. 8 for the non-wrapping situation.

Unlike Fig. 6, there is no sharp change in conditions for queries at half the region of interest size for a non-wrapping world. At half the world, there is just probability 1 that the query sphere will intersect the outside region, with more and more cell offsets falling outside as distance grows. This possibility was present for lower and lower distances too, albeit with a decreasing probability less than 1.

Fig. 8 exposes a similar view as Fig. 1 for a non-cyclic world. The gains are generally lower for the non-cyclic case due to the boundary effects. Due to the aforementioned volume estimation procedure, and the bin-lattice not considering the whole world, the proposed algorithm

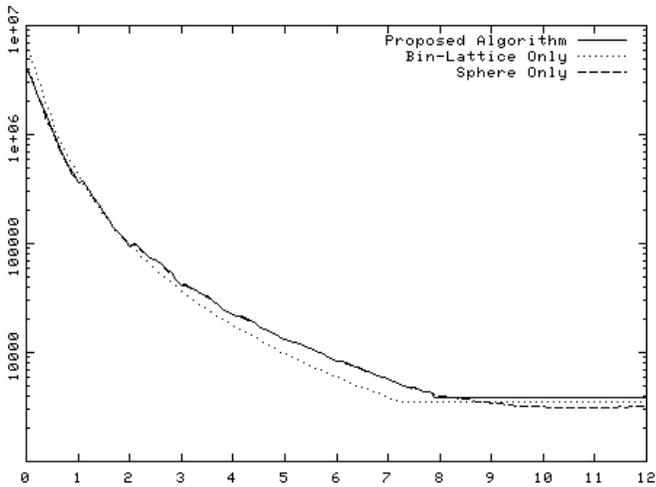
Fig. 7. Performances (queries/s) vs query distance (cell units), for finding all neighbors within range, in a 16x16x16 wrapping world, with a load average of 1 object/cell.

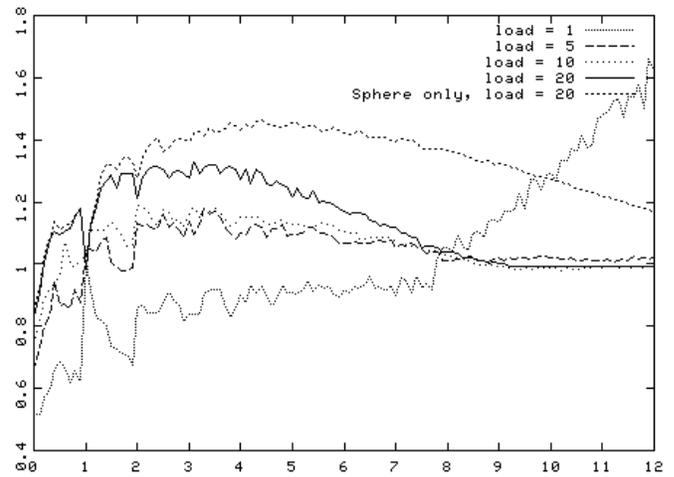
Fig. 8. Proposed algorithm / bin-lattice performance ratio vs query distance (cell units), for a 16x16x16 non-wrapping world, for different cell load averages measured in objects/cell.

smoothly converges to the bin-lattice one without a sharp drop for loads of 5, 10 and 20. For a low load average of 1, the non-empty list is used. But unlike the situation in Fig. 6, Fig. 7 has shown that the base bin-lattice performances continue to decrease after half the world size. Therefore, in Fig. 8, it is logical to observe a performance ratio increase, since the non-empty list performance remains nearly constant whatever the query distance (it just includes more and more calls to the user-provided functor as distance grows). However, there is a non-continuous jump in performance ratios when switching to the non-empty list. There is no smooth blending of methods like the sphere/bin-lattice transition that was presented for Fig. 7: the non-empty list presents a constant volume, unlike the bin-lattice near half the world size. Moreover, the processing and memory access costs are different in the non-empty list case. For example, the sphere method could unconditionally include some objects, but the non-empty list method cannot do so with the current implementation[1].

An additional observation is plotted in Fig. 8.: the ratio between the performance of the spherical offset method only, without volume comparison to select another method, with the bin-lattice performance. For the same volume, any cell that falls outside the region of interest is empty for the sphere method. For the same distance, the sphere only algorithm may unconditionally include some cells entirely within the query sphere. Both these effects are amplified by a high load average, 20 in this case. In such situations, the volume comparison is therefore a bad indicator of the true performances of the algorithm: The sphere only method may become faster even if it processes a greater number of cells. The effect is even visible for small distances, though the gain is less dramatic since the probability to reach the outside region is smaller and there are less interior cells.

To moderate that observation, Fig. 7 shows that the sphere only method brings lesser improvements for a lower load average of 10 for short distances, an no improvement for large distances. Details why weighting the volume comparison may be beneficial to some applications are exposed in Appendix. The reference implementation has this weighting possibility built in, since it can greatly improve performances in some cases. The default is to compare only the volumes as this is the best indicator in most cases. All benchmarks presented in this document select the method to use based only on the volume comparison. A real application would benefit from experimenting with the weighting parameter.

Fig. 9 introduces the influence of the discretization size. All other parameters are kept the same, including the position in space of the query centers and of the objects. These plots suggest there exists an optimal discretization size.

With the finer-grained discretization (32x32x32 case) more cells need to be processed, but globally less objects outside range are present in these cells. The result of this trade-off depends on the total number of objects. The optimal discretization in the case of the present experiment could possibly be non-power of two. But then, the cost that would be introduced by such discretization values could be higher than the gain.

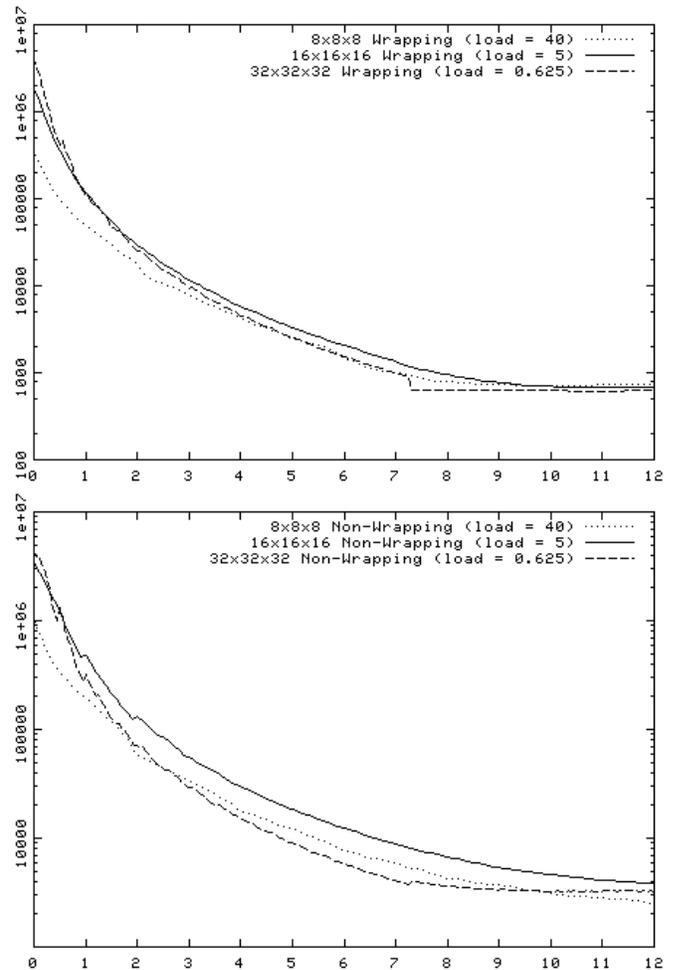

Fig. 9. Influence of the discretization size, when the number of objects is kept constant. Performances are plotted for distances expressed in units of 16x16x16 world cells size. The top figure is the case for a wrapping world, the bottom for a non-cyclic one. Loads are average number of objects/cell.

There is little qualitative difference between the cyclic and non-cyclic versions. The aforementioned non-empty list switch effect is visible on both graphs for the 0.625 load curve, though the performance jump in each case is reversed. The smaller cell size has a better beneficial effect in the wrapping world case, where there is no boundary effect so all the finer cells are inside the region of interest.

### B. Finding only the closest neighbors

All aforementioned influential factors are of course still present when the algorithm is asked to find only the closest neighbors instead of all neighbors within range.

The notable exception is the volume comparison selector: Since the algorithm will terminate earlier when the desired number of neighbors is reached, by reducing the maximum distance to search for other candidates, it is not known in advance how many cells will be processed. Hence the initial sphere estimated volume is not applicable. However, this is compensated by the early cut-off, which makes such a volume test irrelevant in most cases anyway.

---

1. For completeness, tests were made to apply a similar check for the non-empty list method as is done for the sphere method: determining whether the non-empty cell falls entirely within the query sphere or not. It was found that for common load averages, the cost of this extra test is not worth the gain it brings, at least with the current implementation. The same is true for the bin-lattice algorithm, where such an additional test is possible too.

Fig. 10 shows the early cut-off in action. Unlike the situation in Fig.9, the best performances are obtained in the case with the finer discretization. This is logical, since the aforementioned drawback of the finer discretization is greater for large distances, and the advantage more effective for short distances (this is also visible on Fig. 9). Note that the cutoff distance is about one cell size in all presented cases, the plot is scaled in 16x16x16 world cell units for comparison purposes.

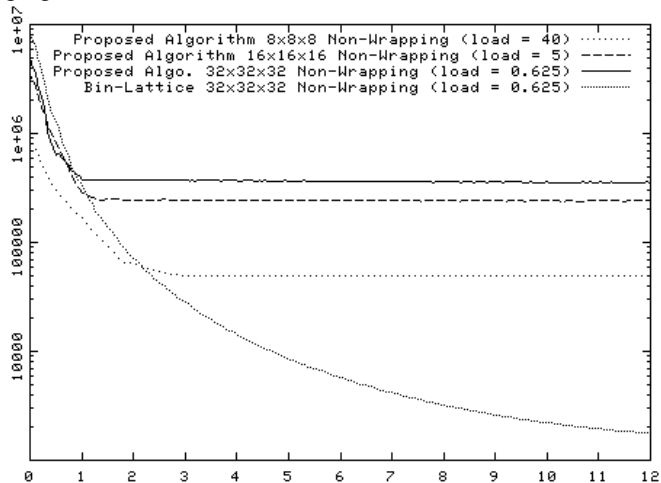

Fig. 10. This legend of this plot is similar to Fig. 9, but for finding only the nearest neighbor. The bin-lattice algorithm was added for comparison.

The cases for finding the K-nearest neighbors instead of just the nearest one are similar to this plot. The only difference is in lower absolute performance values. Since there are 26 cells around the center cell, for K < 27 * load, the cutoff distances are similar to this plot.

The wrapping case is not shown; it is qualitatively not different enough to justify an inclusion in this document. Especially because in the wrapping case, the non-empty list and volume considerations are not applicable. Only the boundary effects remain, and these are effective mainly for large distances. The non-wrapping case was chosen to show the worse situation.

In any case, the improvement due to the ability of the sphere offset algorithm to early stop as soon as the K neighbors have been found is dramatic, compared to the bin-lattice algorithm, especially for large distances. For very small distances, the same situation arises as in Fig. 5 and 8., and it may be worth reverting to the bin-lattice depending on the load average. This possibility has been taken into account in the reference implementation, and is detailed in the Appendix.

## V. POSSIBLE EXTENSIONS

One interesting extension would be toward more dimensions. It's a known fact [3] that the ratio between a sphere volume and its bounding cube volume decreases with the dimensionality, already at ~52% for three dimensions, ~31% for four dimensions, etc, and less than 1% in dimensions 9 and above. Hence the gains brought by a spherical indexing scheme should become increasingly important with the dimension. In practice, the algorithm is then limited by the memory consumption necessary to store the tables and cell arrays, which may be acceptable or not depending on the application, or even intractable in large dimensions.

Conversely, in two dimensions, the algorithm is less interesting, with a circle/square surface ratio of ~79%. Nevertheless, for applications where locality queries tend to consume most of the available time, a potential improvement should not be neglected. Moreover, the unconditional inclusion (all neighbors) and early termination (K-nearest neighbors) optimizations are applicable to any dimension.

Another possibility in three dimensions could be to stack such two-dimensional circle indexed planes, so as to reconstruct the query sphere. For applications where most objects are concentrated in a few planes this could bring a significant improvement. For example, agents evolving on a bumpy terrain with a few flying objects. In this case, running through the whole sphere volume seems a waste of time. Stacking two-dimensional versions of the present algorithm would allow to compare the non-empty lists together with the stacked circles making up the query sphere, for each plane. If the agents truly lie on a few selected planes, most of the sphere will be avoided because the non-empty cell lists will be used for most of the planes (and possibly the lists will themselves be empty), while the potential optimization is still retained for the useful planes.

Whatever the use of this algorithm, and the possible future extensions, the mere disposal of a spherical indexing scheme comes with the corresponding potential improvements and optimizations.

## VI. CONCLUSION

Spherical indexing of a discretized space allows to get a single list of all the cells comprising the query sphere, for any radius and center position. Running through this list, implemented as a pre-computed sorted offset array, is very efficient.

The particular cases where the edge boundaries or the small distances do not allow to take full advantage of the spherical algorithm can be easily reverted to the simpler bin-lattice situation. The reference implementation has this capability built-in, though it was disabled in the benchmark tests for obvious analysis reasons. In any case, the provided implementation is faster than the previous bin-lattice one using double-linked lists, with a ratio that is even more important than the algorithmic improvement itself in many cases.

This algorithm would be especially well adapted, compared to the bin-lattice one, to situations including large number of objects, like point clouds. The wrapping worlds that are common in multi-agent simulations would benefit most from this algorithm as well. The high improvement ratios of Fig. 5 are applicable in that case. This algorithm is also well suited for problems like signaling and communication, where all agents in sight must be contacted regardless of their distance.

The K-nearest neighbor finding problem benefits directly from the spherical indexing, with the ability to early stop when the neighbors are found. This algorithm is therefore valuable also in the K-nearest neighbors finding situations.

As usual for any algorithm, the proper usage depends on the application. For static environments with fixed object positions, some other methods like sorted trees may be more efficient. For dynamic situations, found for example in real-time interactive applications, this algorithm may be a good choice.

## Appendix: Reference Implementation

A C++ reference implementation of this algorithm is available, links can be found on the author web page http://nicolas.brodu.free.fr. This implementation is optimized for 64 bytes memory cache lines, though it will work on older machines as well. It was written for simple precision IEEE 754 floating-points (32 bits). It can handle cyclicity along 2 or 3 dimensions, and non-cyclic regions of interest.

The idea behind the spherical indexing algorithm is to avoid processing unnecessary cells. This saves on both the number of memory access, and also the time necessary to run through all the objects in these cells.

But running a triple loop over the bounding cube, one for each dimension, is very simple and can be implemented with little overhead. On the other hand, the spherical indexing algorithm goes through a distance index table, which then gives the offsets of the cells that are accessed, after translation from the query center. The constant-time $c$ necessary to set up the sphere algorithm is larger than in the simple triple-loop case. The per-cell cost $\alpha$ is globally lower for the spherical indexing for high load averages, but higher for small load averages. On the one hand the offset indirection is costly, though not that much: the offset table is a contiguous array so new memory lines need only be accessed every so often and may even be prefetched by the hardware. This indirection cost is independent from the load average. On the other hand objects in cells below $d - \sqrt{3}$ are included unconditionally, which reduces $\alpha$. This reduction is more advantageous for high load averages, but less effective for small distances.

Some other parameters to take into account include the code size of the inner-loop critical section, including the user provided functor to apply to the neighbors, so it ideally would fit in the trace cache. The memory prefetching capabilities of the machine should be considered, and they are usually combined with a hardware branching predictor (which also tends to reduce the cost of loops). Minimizing the number of floating-point operations by detecting bit-patterns is also influential. The memory cache line size itself crucially determines how often a new line is accessed, etc.

The data structures thus highly influence the performance outcome of the algorithm, together with good memory management (especially regarding the usage of cache lines). Some aggressive optimizations were performed, especially bit manipulations to avoid branching, IEEE 754 floating-point representation assumptions, some more parallel computations within a register like those presented in section II, and C++ template partial specialization to generate optimal code. The algorithm was designed to be backward compatible with older machines, so it does not include floating point vector computation extensions, and no assembly, though these would certainly improve performances too, especially for the distance computations.

A weighting factor for the run-time method selection (sphere or parallelepiped) has been introduced. Each method has different constant and per-cell processing costs. The estimated or exact volume that would be processed by each method is thus not entirely representative of that method performance. Moreover, the processing costs may depend on the query distance, like the spherical indexing unconditionally includes all objects below a certain distance. In addition, for non-wrapping worlds, for the same volume some offset cells may fall outside the region of interest, with again different processing costs. The optimal weighting factor may thus be determined experimentally, for some representative conditions of the real application. All the benchmarks presented in the main section use a direct volume comparison.

For small distances an optional reversal to bin-lattice is available, but deactivated by default. The reversal distance is a parameter of the algorithm, that should be set depending on the application. Since in Fig. 5 and 8 some improvements are still observed for high load averages and/or wrapping worlds, with distances below 1 cell, the user is invited to perform tests to determine the best value for a particular application case. Similarly, a distinct distance reversal parameter has been introduced for the K-nearest neighbors algorithm due to different setup costs compared to the all-neighbor queries. That parameter can be optimized to select the best method before the cut-off.